# A 25 Gb/s Silicon Photonics Platform

Tom Baehr-Jones[1,*], Ran Ding[1], Ali Ayazi[1], Thierry Pinguet[1], Matt Streshinsky[1], Nick Harris[1], Jing Li[1], Li He[1], Mike Gould[1], Yi Zhang[1], Andy Eu-Jin Lim[2], Tsung-Yang Liow[2], Selin Hwee-Gee Teo[2], Guo-Qiang Lo[2], and Michael Hochberg[3]

[1]*Department of Electrical Engineering, University of Washington, Campus Box 352500, Seattle, Washington 98195, USA*
[2]*Institute of Microelectronics, Agency for Science, Technology and Research (A*STAR), Singapore 117685, Singapore*
[3]*Department of Electrical and Computer Engineering, University of Delaware, 140 Evans Hall, Newark, DE 19716, USA*
[*]*baehrjt@washington.edu*

**Abstract:** Silicon has attracted attention as an inexpensive and scalable material system for photonic-electronic, system-on-chip development. For this, a platform with both photodetectors and modulators working at high speeds, with excellent cross-wafer uniformity, is needed. We demonstrate an optical-lithography, wafer-scale photonics platform with 25 Gb/s operation. We also demonstrate modulation with an ultra-low drive voltage of 1 Vpp at 25 Gb/s. We demonstrate attractive cross-wafer uniformity, and provide detailed information about the device geometry. Our platform is available to the community as part of a photonics shuttle service.

## 1. Introduction

Silicon has attracted attention as an inexpensive and scalable material system for photonic-electronic [1,2] system-on-chip development. For this, a platform with both photodetectors and modulators working at high speeds, with excellent cross-wafer uniformity, is needed [3-5]. The ideal platform will also be based on optical lithography for wafer-scale fabrication, and demonstrate high speed, low cross-wafer variability and low defectivity rates. Here we report a new wafer-scale silicon photonics platform with integrated modulators and photodetectors demonstrating 15 GHz bandwidths and higher, as well as 25 Gb/s operation. We demonstrate modulation with an ultra-low drive voltage of 1 Vpp at 25 Gb/s, the first published example of a 25 Gb/s modulator directly compatible with typical CMOS transistor voltages. We demonstrate attractive cross-wafer uniformity, and provide detailed information about the device geometry. Our platform is available to the community as part of a photonics shuttle service [6-8].

The first multi-Gb/s capable silicon photonics modulator was reported by Liu et al in 2004 [9]. Since then, silicon photonics has seen an explosion of interest, with low-loss waveguides [10], detectors [11], and grating couplers [12] all separately demonstrated. A challenge has been the development of a platform that supports all of these key components simultaneously with attractive performance, and achieves drive voltages compatible with CMOS processes. Luxtera demonstrated a photonics platform with both integrated modulators and detectors at 10 Gb/s in 2006 [13]. Luxtera has since announced 25 Gb/s capability [4] and recently reported a 4x28G link [14], though the technical details remain unpublished, and key device geometries and performance characteristics have been kept a trade secret. Kotura has achieved 40 Gb/s for various devices [15], but has not reported a common platform with both high-speed modulators and detectors on the same wafer, at the same time. Intel and IBM have reported individual high-performance devices [16,17]. Meanwhile, ePIXfab has been providing fabrication services to the photonics community [18], and has published results on modulators and detectors at 40 Gb/s speeds separately [19,20], though not on results achieved simultaneously as part of an integrated platform. Kim et al have recently demonstrated both modulators and detectors working at speeds of 30 Gb/s [5]; however, the on-chip drive voltage required for the modulator was 2.4 Vpp, and additionally, a 5 V DC bias was required for 15 GHz+ bandwidths, which will hamper integration with advanced CMOS circuits. Wafer-scale defectivity and performance data was also not provided for any of the devices in question, making it unclear how suitable this platform will be for large scale photonic integration.

Silicon is not the only viable integrated photonics platform. For example, Infinera has been pursuing photonics integration in III-V systems, with 45.6 Gb/s speeds achieved [21]. Devices in III-V systems have traditionally enjoyed performance advantages compared to silicon. Traditional semiconductor light sources are impossible in silicon, while III-V lasers are in wide use, and III-V based modulators enjoy better performance [22] than state-of-the-art silicon devices. This is similar to the electronics industry; typically non-silicon platforms have superior performance. Silicon has become dominant due to low cost and high yield, which have enabled integrated systems of unparalleled complexity to be constructed [23]. To realize these benefits in the photonics community, a high-speed and low-defectivity wafer-scale platform is necessary.

## 2. Fabrication

Fabrication for this work occurred at the Institute of Microelectronics (IME)/ASTAR [8]. The starting material was an 8" Silicon-on-Insulator (SOI) wafer from SOITEC, with a Boron-doped top silicon layer of around 10 ohm-cm resistivity and 220 nm thickness, a 2 μm bottom oxide thickness, and a 750 ohm-cm handle silicon wafer, needed for RF performance [24]. A 60 nm anisotropic dry etch was first applied to form the trenches of the grating couplers. Next, the rib and channel waveguides were formed using additional etch steps. In all cases 248 nm photolithography was utilized. The p++, p, n++, and n implants for the Si modulator and the p-type doping for anode formation of the Ge p-i-n photodetectors were performed on the exposed silicon, prior to any oxide fill. This was followed by a rapid thermal anneal at 1030 ºC for 5 s for Si dopant activation. Ge was then selectively grown to a thickness of 500 nm [25] in regions defined by a deposited $SiO_2$ mask layer. Ion implantation was performed for the Ge regions and annealed at 500 ºC for 5 min, followed by the formation of contact vias and two levels of aluminum interconnects. Chemical-mechanical planarization (CMP) was not utilized. The schematic cross-section is shown in figure 1. Excluding edge dies, 41 dies of overall size 25 x 16 mm were fabricated per wafer. All results reported are from a single wafer.

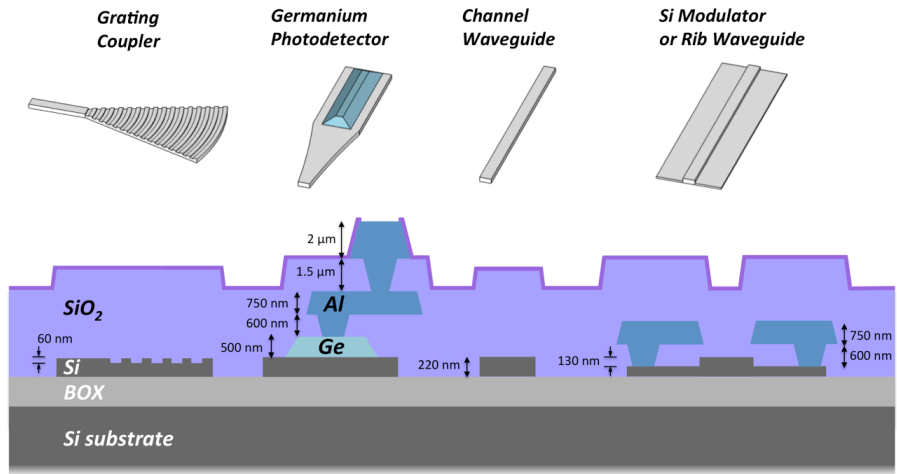

Fig. 1. Platform geometry. A diagram of the cross-sections of key devices is shown, as well as layer thicknesses. The platform features a 220 nm top silicon layer that has three possible etch depths, an epitaxially grown Ge layer, and two metal layers for electronic routing.

Implant recipes were chosen to give peak doping concentrations of approximately $7 \times 10^{17}$ cm$^{-3}$ for the p doping regions, $1.7 \times 10^{20}$ cm$^{-3}$ for the p++ doping region, $5 \times 10^{17}$ cm$^{-3}$ for the n doping concentration, and $5 \times 10^{20}$ cm$^{-3}$ for the n++ doping concentration. The sheet resistances were measured to be 10 kΩ/□, 110 Ω/□, 7 kΩ/□, and 58 Ω/□, respectively. These values are for the resistance in the partially etched silicon regions.

## 3. Devices and Results

*3.1 Waveguides*

All test results reported here were obtained using a wafer-scale, normal incidence opto-electronic test setup. Fiber-coupling was achieved with grating couplers [26]. The average grating coupler insertion loss across 19 dies was determined to be 4.4±0.2 dB, with a peak wavelength near 1545 nm and a typical 1.5 dB bandwidth of 45 nm. The typical on-chip return loss from the grating couplers was -12.5 dB. The defectivity rate for characterization loops was 5%, out of 657 devices tested (see section 4 for further details). The period and size used for the trenches was 630 nm and 220 nm, respectively. Cross-wafer data is shown in figure 2. Waveguide losses of 1.0±0.3 dB/cm were achieved in the best die measured for 0.5 µm wide rib waveguides with 220 nm thickness, with low cross-wafer variation as shown in figure 2. The average waveguide loss was 2.2±0.8 dB/cm from 19 dies measured. For channel waveguides with 0.5 µm dimensions, losses of 2.4±0.3 dB/cm were obtained from these dies. On the worst-performing dies, losses were 3.4±0.2 dB/cm for the rib waveguide, while worst-case losses of 3.1±0.1 dB/cm were seen for the channel waveguide.

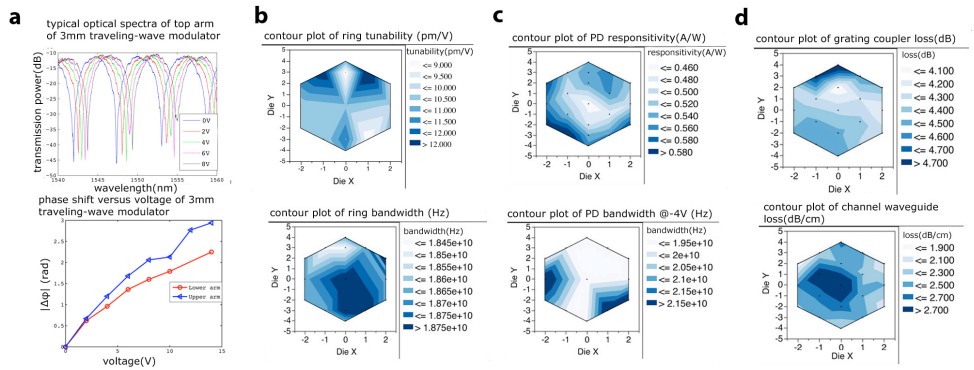

Fig. 2. Cross-wafer data. (a) The performance of the traveling-wave Mach-Zehnder is shown. Typical spectra with varying applied reverse-bias voltage is shown, as well as phase shift as a function of bias voltage. (b) The performance of the ring modulator across the wafer is shown, with both tunability and bandwidth reported. (c) Ge Photodetector responsivity is also shown, as well as bandwidth with a 4 V reverse-bias. (d) Grating coupler performance is shown, as well as the cross-wafer variation in channel waveguide loss.

*3.2 Traveling-wave Modulators*

Traveling-wave modulators were constructed [27, 28], based on a lateral pn-junction waveguide [24] with coplanar metal strips, with a GS configuration utilized. The device length was 3 mm, while each arm had an independent GS transmission line drive. Each GS line was predicted to have an impedance of around 33 Ω, consistent with the measured S11 values. A lower impedance assists velocity-matching and lowers RF losses [24]. Termination was achieved with two 50Ω lines, via a GSSG probe. As detailed in section 4, a test pattern of 9 dies was used. Figure 3 shows the device layout. The average cross-wafer modulator insertion loss, excluding known losses from the grating coupler and routing waveguides, was 6.2±0.5 dB. The device was intentionally unbalanced by a waveguide length of 100 μm, to enable biasing of the modulator by tuning the input wavelength. A slight asymmetry in response was observed between the two arms. We characterize the voltage required for a phase shift of π/2 radians, due to the well-known nonlinear electrooptic response of reverse-biased pn junctions, as shown in figure 2. The cross-wafer average Vπ/2 value for the lower arm was 7.2±1.3 V and 5.3±0.2 V for the upper arm.

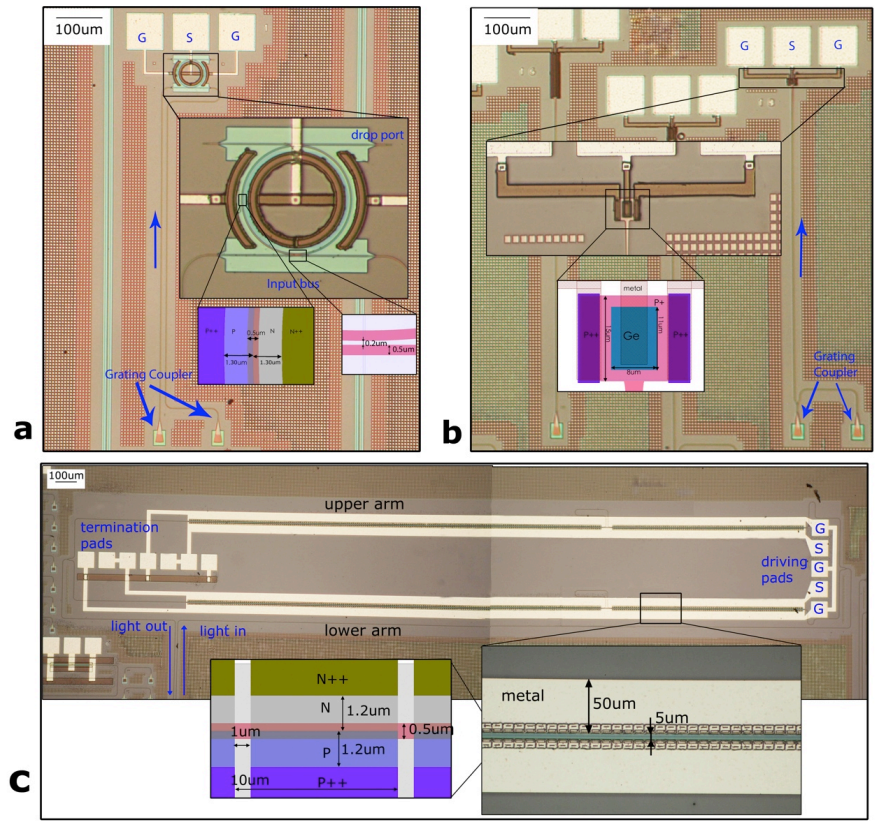

Fig. 3. Device layout. (a) Optical micrographs of the ring modulator device. Note that the insets with the highest magnification are renderings of the mask layout, instead of micrographs. (b) Optical micrographs of the Ge detector device. Again, the highest magnification insets are renderings of the mask

layout. (c) Optical micrographs of the 3 mm traveling-wave Mach-Zehnder modulator.

Average cross-wafer 3 dB bandwidths at 0 V bias were 15.5±0.2 GHz for the lower arm, and 17.6±0.8 GHz for the upper arm. The cross-wafer variation of some key parameters is shown in figure 2. Eye-patterns were taken with differential drive at 25 Gb/s, utilizing a pseudo-random bit sequence. An extinction of 4.4 dB was obtained with a 1 Vpp signal, as shown in figure 4. The drive signal was centered at 0 V. We note that even for small forward biases, forward diode current remained minimal due to the built-in voltage of the junction. For this particular Mach-Zehnder device, a lower arm V$\pi$/2 of 6.8 V and upper arm V$\pi$/2 of 5.5 V was measured. At 0 V bias, the typical capacitance of the lateral pn junction was found to be 0.3 fF/μm. For the typical separations of 1.05 μm between the n++ and p++ doping and the unetched ridge, intrinsic junction bandwidths of 31 GHz should be possible based on measurements of the relevant sheet resistances. Because the modulator had impedance lower than 50 Ω, the true on-chip voltage was in fact slightly lower than 1 Vpp.

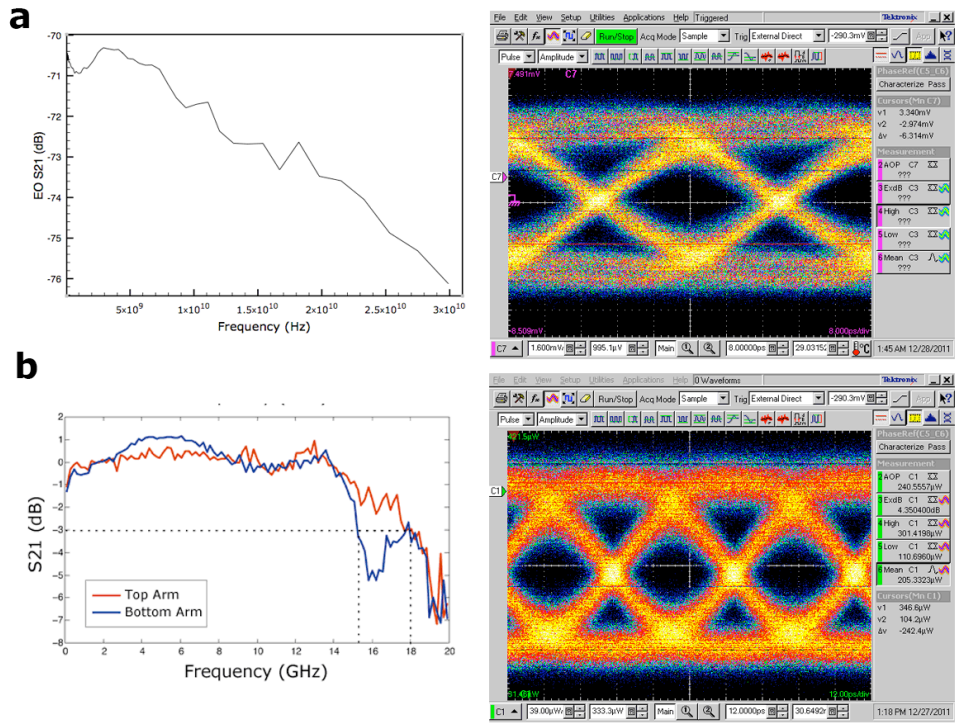

Fig. 4. Traveling-wave Modulator and Detector Performance. (a) The electrooptic (EO) S21 of a typical photodetector with 2 V reverse-bias, as well as an eye pattern at 25 Gb/s. The 3 dB rolloff is at 20 GHz. (b) The EO S21 traces of a typical traveling-wave Mach-Zehnder device are shown, as well as an eye pattern produced by a differential drive signal of 1 Vpp at 25 Gb/s. 4.4 dB of extinction is achieved. 5.5 dB of excess loss in addition to the intrinsic device loss is seen on the modulator for a "1" bit, due to the device bias point.

*3.3 Ring Modulators*

Dual-bus ring modulators [29,30], with a 30 µm radius and a lateral PN junction [31] centered in a rib waveguide were tested in an expanded set of 13 dies (see later section on wafer test patterns). The ring FSR was around 3.2 nm, and a typical Q was 7,000. An average cross-wafer tunability of 10.6±1.2 pm/V was determined by measuring resonance shift as bias voltage varied from -200 mV to 200 mV. Ring bandwidth was measured at 0 V dc bias. The average bandwidth was 18.7±0.1 GHz, as shown in figure 2. No defective ring modulators or photodetectors were encountered during test.

*3.4 Photodetectors*

Photodetectors were constructed with vertical Germanium PIN junctions, as shown in figure 1. Devices were 11 µm long, 8 µm wide, as shown in figure 3. Average on-chip responsivities of 0.54±0.05 A/W with dark currents of 4.8±0.4 µA were achieved at 4 V reverse-bias. RF testing was also performed, with devices showing average cross-wafer 3 dB bandwidths of 20.2±1.4 GHz, indicated in figures 2 and 4. The bandwidth was determined from the S21 parameter of a vector-network analyzer driving an external modulator. Several devices were tested at lower reverse-bias voltage as well; one detector exhibited bandwidth of 20 GHz with 2 V reverse-bias, as shown in figure 4. A dark current of 1.2 µA and responsivity of 0.46 A/W was measured. One detector was characterized for nonlinearity and the damage threshold, which was 15 mW of on-chip optical power. The detector exhibited a 14% deviation from linear responsivity at 7.8 mW input power. An eye pattern is shown at 25 Gb/s in figure 4. Typical detector capacitance was measured at 4 V reverse-bias to be 0.5 fF/µm$^2$, or 44 fF.

*3.5 Testing temperatures*

We briefly mention the temperatures at which we perform measurements. Passive characterization on waveguides and grating couplers was performed at room temperature. Characterization of the active components, such as photodetectors and modulators, was done at 35 °C. The eye patterns were taken at 30 °C. This temperature was chosen as the thermal chuck used in testing was observed to be maximally stable at this value. However, no significant temperature dependent changes in the performance of any component seen – aside from the well-known shift in ring resonance peak location.

*3.6 RF Testing*

The photodetector and modulator devices were probed with standard GSG RF probes. The former required a bias tee to introduce a DC bias into the system. Driving the traveling-wave devices was achieved through a standard 50 Ω GSG probe to introduce an RF signal on to the chip, and a terminating RF probe with GSSG configuration. Two 50 Ω lines in parallel achieved a matching 25 Ω termination, close to the 33 Ω line. Because the S11 coefficient for a 50 Ω line driving a 33Ω load is nearly -14 dB despite the mismatch, it was possible to drive the traveling wave device directly for the eye pattern measurement. For the eye pattern, GSGSG and GSSGSSG probes were used, as both arms were driven differentially. The standard set of RF calibrations were used for the S21 values shown. Additionally, a slight frequency nonuniformity in a commercial lithium niobate modulator used in some testing was calibrated out. For the electrooptic S21 parameter of the modulators, an external high-bandwidth photodetector was used to convert the modulated optical signal back into RF input for a vector network analyzer.

## 4. Cross-Wafer Testing

Here we briefly expand on the testing methodology used. A subset of dies was tested. Figure 5 shows the testing pattern used for grating couplers. Note that the edge dies are also shown in the figure; edge dies were not tested as in all cases, as their proximity to the edge of the wafer suggested they would have uncharacteristically large levels of defectivity or other variation.

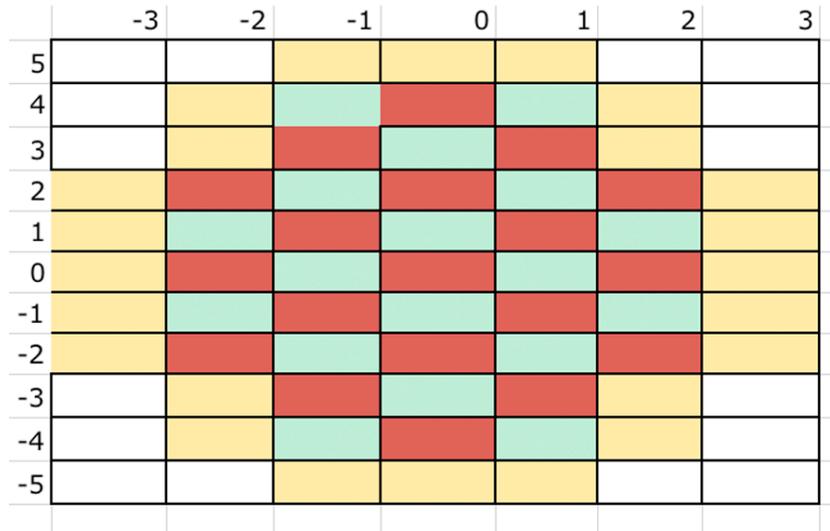

Fig. 5. Die pattern used in testing for grating coupler and waveguide loss measurements. The red indices indicate that testing was performed on the specified die, while green squares indicate potentially good dies that were not tested. Yellow indicates an edge die. The die x (horizontal), y (vertical) indices indicate the relative position on the wafer.

During our test, some grating coupler calibration loops returned dramatically less light than average. Out of 657 calibration loops tested across all 19 dies, a total of 30 loops exhibited more than 4 dB excess loss as compared to the average on each die, amounting to a defectivity rate of 5%. Typically only 1 or 2 calibration loops had this level of excess loss on each die, with 5 calibration loops exhibiting this level of loss in the worst case. Testing was not performed in a cleanroom environment, and a visually noticeable amount of dust appeared on the wafer during the testing process. In a few cases, devices that did not return light began to function after the fiber array on the test setup was cleaned. The 30 loops that were defective were not re-tested. It is conceivable that the true defectivity rate is lower than what is reported here. Similar testing patterns were used for all other measurements, though it was not practical to test as many dies for all other types of devices.

For traveling-wave modulator characterization, 9 dies were characterized, as shown in figure 6.

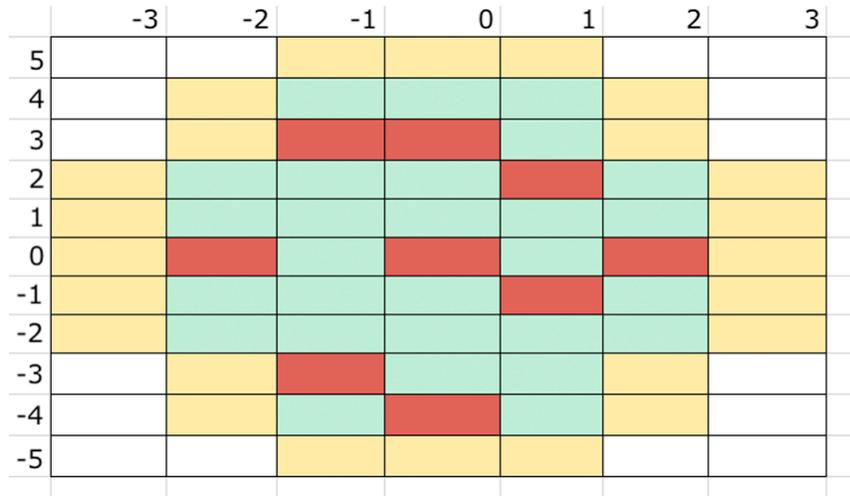

Fig. 6. Die pattern used in testing for traveling-wave characterization.

We do not report a defectivity rate for the traveling-wave devices. Testing these devices was challenging, due to the multiple RF contacts needed. Pad erosion during test was a significant issue, and in a few cases devices were destroyed before they could be fully characterized. The mechanism of destruction was always metal from adjacent RF pads causing a short when the probes were in contact with the pads, which was confirmed by visual inspection and resistance measurements. In all cases, what data was measured agreed with the results presented for the fully characterized dies. We suspect the traveling-wave defectivity rate is quite low, however wire-bonding or bump-bonding will be likely required to prove this, as these may be more reliable ways of obtaining good RF contact.

The die pattern used for the ring modulators and photodetectors is presented in figure 7. In a few cases, additional data was taken from additional dies for some key parameters, such as dark current. For simplicity, the numbers presented in the main paper are solely from the 13 dies fully characterized. Figure 3, however, does include all available data points. All measurements taken were in statistical agreement with the cross-wafer averages, and no defective devices were encountered.

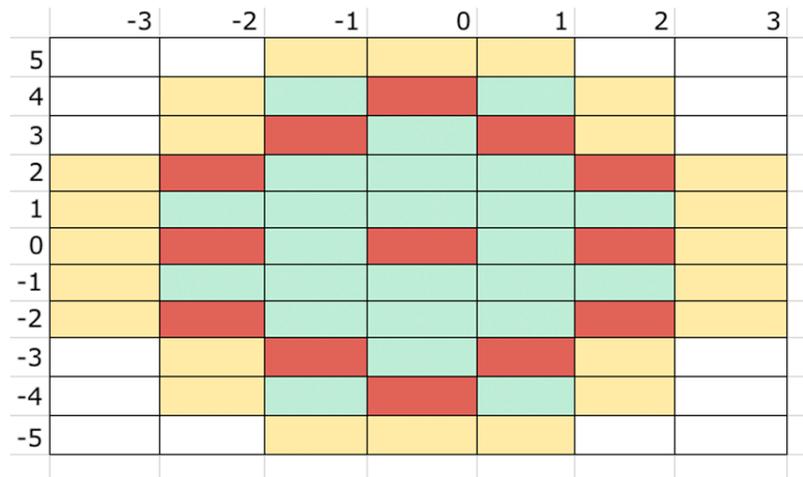

Fig. 7. Die pattern used in testing photodetector and ring modulator characterization.

## 5. Traveling-Wave Modulator Layout

The layout of the traveling-wave modulator is shown in increased detail in figures 8-9. The striations seen in the p and n regions are designed to ensure that the majority of the current flows in the metal transmission line and not the silicon, to ensure that RF losses remain low. The pn junction for the lower arm is oriented differently than the upper arm; the n-type silicon is on the internal side of the modulator in both cases. This may account for the asymmetry in response seen. If there were an overlay error in the vertical direction with one or both of the implants, it could explain the observed behavior, since the two junctions would no longer be symmetric.

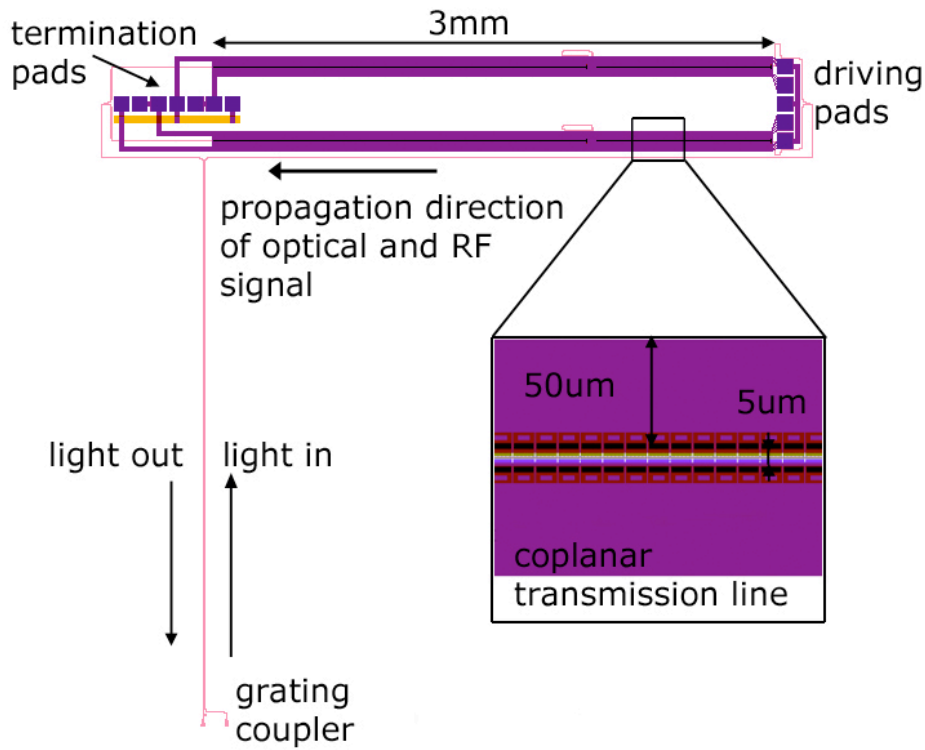

Fig. 8. Die pattern used in testing photodetector and ring modulator characterization. The large clearance from the device to the grating couplers enables simultaneous optical and electrical probing.

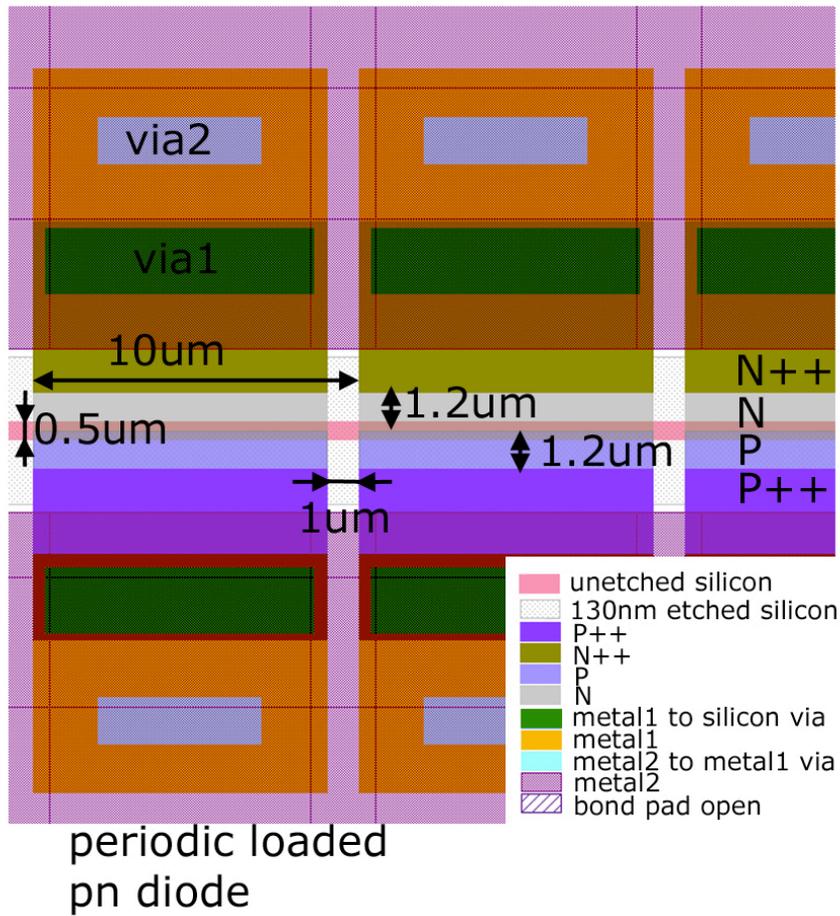

Fig. 9. Die pattern used in testing photodetector and ring modulator characterization.

## 6. Conclusions

To conclude, we have demonstrated a wafer-scale silicon photonics platform that achieves 15 GHz bandwidths, and is capable of 25 Gb/s data rates. We provide precise fabrication details and device geometries, which should enable other users of this platform to fully leverage our results in designing integrated photonic-electronic systems-on-chip, and in developing their own fully integrated processes.

**Acknowledgements**

Assistance with 25 Gb/s testing from Klaus Engenhardt at Tektronix is gratefully acknowledged. The authors would like to thank Gernot Pomrenke, of the Air Force Office of Scientific Research, for his support under the OPSIS and PECASE programs, and would like to thank Mario Panniccia and Justin Rattner, of Intel, for their support of the Institute for Photonic Integration.